\documentclass[aip,cha,superscriptaddress,twocolumn,numerical author-year,reprint]{revtex4}

\usepackage{amsfonts,amssymb,amsmath,wasysym,times}
\usepackage{color}
\usepackage{algorithm}
\usepackage{algpseudocode}
\usepackage{graphicx} 

\usepackage[usenames,dvipsnames]{xcolor}

\usepackage{url}
\usepackage[breaklinks]{hyperref}
\usepackage{breakurl}

\begin{document}

\title{Fluctuation of similarity (FLUS) to detect transitions between distinct dynamical regimes in  short time series}

\author{Nishant Malik}
\affiliation{Department of Mathematics, CB \#3250, University of North Carolina - Chapel Hill, NC 27599, USA} 

\author{Norbert Marwan}
 \affiliation{Potsdam Institute for Climate Impact Research, P.\,O.~Box 601203, 14412 Potsdam, Germany}

\author{Yong Zou}
\affiliation{Department of Physics, East China Normal University - Shanghai 200241, China }
\affiliation{Potsdam Institute for Climate Impact Research, P.\,O.~Box 601203, 14412 Potsdam, Germany}   
    
\author{Peter J. Mucha} 
\affiliation{Department of Mathematics, CB \#3250, University of North Carolina - Chapel Hill, NC 27599, USA} 

\author{J\"urgen Kurths}
   \affiliation{Potsdam Institute for Climate Impact Research, P.\,O.~Box 601203, 14412 Potsdam, Germany}
   \affiliation{Department of Physics, Humboldt University Berlin, Newtonstr.~15, 12489 Berlin, Germany}

\date{\today}

\begin{abstract}

Recently a method which employs computing of fluctuations in a measure of nonlinear similarity  based on local recurrence properties in a univariate time series, was introduced to identify distinct dynamical regimes and transitions between them in a short time series \cite{nmalik_epl1}. Here we  present the details of the analytical relationships between the newly introduced measure and the well known concepts of attractor dimensions and Lyapunov  exponents. We show that the new measure has linear dependence on the effective dimension of the attractor and it measures the variations in the sum of the Lyapunov  spectrum. To illustrate the practical usefulness of the method, we employ it to identify various types of dynamical transitions in different nonlinear models. Also, we present testbed examples for the new method's robustness against the presence of noise and missing values in the time series. Furthermore, we use this method to analyze time series from the field of social dynamics, where we present an analysis of the US crime record's time series from the year 1975 to 1993. Using this method, we have found that dynamical complexity in robberies was influenced by the unemployment rate till late 1980's. We have also observed a dynamical transition in homicide and robbery rates in the late 1980's and early 1990's, leading to increase in the dynamical complexity of these rates.      
    
\end{abstract}
 
\maketitle

\section{Introduction }\label{sec:sec0}

One of the central challenges in nonlinear time series analysis has been to develop methodologies to identify and predict dynamical transitions, i.e., time points where the dynamics show a qualitative change \cite{nmalik_epl1,norbert_phyrep, holger,small-bk,hab1,schreiber1997,trans1,lnmeth3,trans2,trans3,lenton,ns-1}. Application of such methods is  widespread in a variety of areas of science and society \cite{scheffer}. For instance, in medical sciences such approaches could be useful in identifying pathological activities of vital organs such as the heart and the brain from ECG and EEG data sets \cite{lehnertz1998,lfs1,lfs2}. Similarly, in earth sciences one can use these  methods to identify tipping elements from modern and paleoclimate data sets \cite{lenton,ashwin,norbert_phyrep,trans1,trans2,trans3}. Also, in the analysis of financial data these methods can be used to better comprehend the behaviour of markets and their vulnerabilities \cite{ec1,econ_1,econ_2}. Apart from these applications, such methods could also be used in the analysis of the evolution of social and economic indictors to understand the well being of a society and to predict probable future changes and also in  physics  to  study the response of an interacting many-body system to an external perturbation \cite{scheffer,ec1,econ_1,econ_2,Chakrabarti_rmp1999}.

What makes this challenge hard is that in a dynamical system there are a variety of reasons which can lead to different levels of qualitative changes in the dynamics of the system \cite{holger,small-bk,ottbook,hab1,ruelle_1,dbif_3}. Some of the most common reasons are the evolving control parameters of the system passing through a bifurcation point, rate of change of these control parameters, internal feedbacks, and noise induced effects \cite{holger,small-bk,ottbook,hab1,dbif_3}. In many natural systems it has been suggested that dynamic bifurcations lead to critical transitions in their dynamical state \cite{dbif_1,dbif_2,dbif_3}. In some cases these transitions are visually more apparent and can be identified with little effort but in some other cases these transitions are much more subtle, especially where transition occurs from one chaotic regime to other complex chaotic regime. For example, in palaeoclimate Dansgaard-Oeschger events on millennial time scales are visible in ice records to the naked eye and have been hypothesised to be caused by a noise induced transition \cite{pdd1,pdd2,stoc-res1}. In contrast, on similar time scales we do not  observe such visibly apparent transitions in many other components of climate, such as the Indian summer monsoon, though it has also gone through dynamical transitions between distinct chaotic regimes due to variations of Milankovitch cycles \cite{mos1,nmalik_epl1,nmarwan-01}. In this case we need more careful analysis.  Similarly in neuroscience, certain brain states like sleep cycling or epileptic seizure are easily detectable from EEG data sets but gamma rhythms or the ultra-slow BOLD rhythms are harder to detect. Again, we need to employ more sophisticated mathematical tools to identify such dynamical states \cite{neuro1}. In our understanding, the intricacies and diversities involved in the origin of dynamical transitions makes it difficult to develop one single method to identify and quantify all possible types of transitions. Rather we need to have a toolbox consisting of several methodologies and approaches inspired from the paradigm of nonlinear dynamics to solve such problems. The case we will be mostly interested in here is the one where the changes in one single control parameter takes the system from a regime of one dynamical complexity to other dynamics of less or higher  complexity, with an important constraint that time series available for the analysis are relatively short (ranging between several hundred to few thousand time points). 

Most widely used methods for some of the above mentioned problems are linear such as auto correlation function and detrended fluctuation analysis  etc. \cite{lnmeth4,lnmeth1,lnmeth2,lnmeth3,lenton}. But certain methods for the analysis of time series using the paradigm of nonlinear dynamics have also shown tremendous promise. Significant among them are the recurrence plot based methodologies such as the recurrence quantification analysis and the recurrence network analysis \cite{norbert_phyrep,Koji-01,Koji-02,Koji-03,trans1,trans2,trans3,yong_power1,donner_1,donner_2}. The method discussed here is called FLUS (FLUctuation of Similarity) in short, and it is based on the concept of nonlinear \emph{similarity} between two time points. It was recently introduced in order to study short paleoclimatic time series of the Indian summer monsoon \cite{nmalik_epl1}. This new method is computationally simple, more automatized, and yet extremely robust in distinguishing distinct dynamical regimes  and in identifying time points where transitions occur between these distinct dynamical regimes,  even in the case where available time series is short. This method also tends to work well in the presence of noise and missing values. In this paper we present analytical findings which relates the new measure to more classical concepts in nonlinear time series analysis such as attractor dimensions and Lyapunov  exponents. To demonstrate the strengths of this method in distinguishing dynamical regimes and in identifying transitions between them, we present a new set of challenging numerical tests and examples of dynamical transition in  different nonlinear models. We also include tests for the new method's robustness against noise and missing values. 

This paper is organized as follows: first we describe the method and some of the analytical results on it with supporting numerics. Then we illustrate the strengths and practical usefulness of this method using several different numerical cases of dynamical transitions in nonlinear systems. Also, we test the method's robustness against the presence of noise and missing values in a pragmatic nonlinear model. This is followed by an application in social dynamics, where we attempt to understand the role of unemployment in the crimes related to robberies and homicides in the US over the period 1975-1993.

\section{Method}\label{sec:sec1}

Let ${\bf{x}}_{j}$ represent the $j$-th vector of a delay embedded
time series of length $N$. The embedding dimension
$m$ and time delay $L$ are estimated respectively by fixed nearest
neighbours and mutual information, as often done in nonlinear time
series analysis ~\cite{holger,small-bk,hab1,phar_1,phar_2}. In this reconstructed phase space we denote the
neighbourhood of any point ${\bf{x}}_{j}$ as $ U
({\bf{x}}_{j})$ containing $k$ nearest neighbors, namely {$U({\bf{x}}_{j}) = \{ {\bf{x}}_{l} : \| {\bf{x}}_{j}-{\bf{x}}_{l} \| < \epsilon_j \}$}, where the set $l$ contains indices of the $k$ nearest neighbours and $\| \cdot \|$ is a norm.  A fixed number of $k$ close-neighbours is chosen for a point ${\bf{x}}_j$, hence $\epsilon_j$ varies with  the change in the values of $k$ i.e., $\epsilon_j=\epsilon_j(k)$. In the text $k$ will be expressed as percentage of total number of points $N$. We use Euclidean distance if not mentioned otherwise.  The point-wise closeness of ${\bf{x}}_j$ to its $k$ neighbours is obtained as the mean distance
\begin{equation} \label{eq:eq3} 
 d({\bf{x}}_{j}) = \frac{1}{k} \sum_{l} \| {\bf{x}}_{j} - {\bf{x}}_{l} \|.
\end{equation}

Next we analyze the evolution of the neighbourhood of ${\bf{x}}_{j}$.
At a later time $j+\tau$, the neighbourhood of
${\bf{x}}_{j+\tau}$ is generally different. But we are mainly interested in the evolution of $U({\bf{x}}_{j})$, i.e., the neighbourhood of ${\bf{x}}_{j}$.Therefore, we calculate the \emph{closeness} of ${\bf{x}}_{j+\tau}$ to the neighbourhood of ${\bf{x}}_j$ by means of a conditional distance, defined 
as
\begin{equation}\label{eq:eq1}
  d({\bf{x}}_{j+\tau} | {\bf{x}}_{j}) = \frac{1}{k}\sum_{l\in U({\bf{x}}_{j}) } \| {\bf{x}}_{j+\tau}-{\bf{x}}_{l+\tau} \|.  
\end{equation}
 The dynamical similarity of ${\bf{x}}_j$ conditioned to
${\bf{x}}_{j+\tau}$  can then be defined by
\begin{equation}\label{eq:eq4}
  S_{j|j+\tau}=\frac{d ( {\bf{x}}_{j}) }{d( {\bf{x}}_{j+\tau} | {\bf{x}}_{j}) }.
\end{equation}

Larger values of $S_{j|j+\tau}$ indicate higher similarities in
the signal (i.e., a periodic trajectory with period $T$, ${\bf{x}}_j =
{\bf{x}}_{j+nT}$ yields a periodic variation of $S_{j|j+\tau}$). It is easy to see that $S_{j|j+\tau}$ is time dependent, relying on
the initial conditions. The distribution of inter-spike interval of
$S_{j|j+\tau}$ reflects the associated recurrent period information, which
shows unique properties for different dynamics (i.e., quasiperiodic or
chaotic~\cite{yong1}). In a full analogy, $S_{j+\tau|j}$ characterising the similarity of
${\bf{x}}_{j+\tau}$ conditioned to ${\bf{x}}_{j}$ can be calculated,
which often yields $S_{j+\tau|j} \neq S_{j|j+\tau}$ since $d( {\bf{x}}_{j+\tau }
| {\bf{x}}_{j}) \neq d( {\bf{x}}_{j} | {\bf{x}}_{j+\tau})$. Previously a similar measure has been used to estimate the nonlinear interdependency between two time series i.e., for bivariate studies ~\cite{arnhold1}, where the conditional distance was calculated between time points coming from two separate time series.

\begin{figure}
 \centering
 \includegraphics [width=\columnwidth]{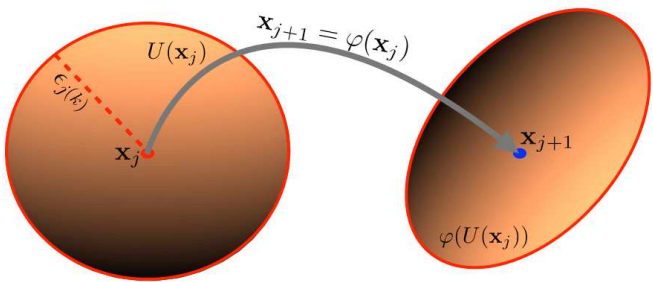}
 \caption{\small { (Color online) Schematic representation of the $\epsilon_j(k)$ ball
     neighbourhood $U({\bf{x}}_{j})$ of ${\bf{x}}_{j}$
     corresponding to $k$-close neighbours of ${\bf{x}}_{j}$ and
     its deformation into an ellipsoid due to the application of
     mapping $\varphi$ on it in a case of ${\bf{x}}_{j+1}=\varphi
     ({\bf{x}}_{j})$ (see region within the red boundary). The
     neighbourhood of ${\bf{x}}_{j+1}$ corresponding to its $k$-nearest
     neighbours is usually different. Locally at ${\bf{x}}_{j}$ the mapping $\varphi$ can be approximated to be a linear transformation, any expansion in the ball $U({\bf{x}}_{j})$ by inclusion of more points will lead to rescaling of the size of $\varphi (U({\bf{x}}_{j})$ by stretching or contraction in different directions. Hence, their radii will scale by the same exponent.}}
    \label{fig:dia1}\end{figure}

$S_{j|j+\tau}$ is a local measure and indicates local properties of the attractor and also it is computationally cumbersome to calculate all the possible $S_{j|j+\tau}$ for a complete time series. Next, we devise a strategy to obtain a measure from $S_{j|j+\tau}$  which is not only computationally simpler, but also has a dependence  on the  global properties of the attractor and hence, it will be sensitive to dynamical transitions. To achieve this task, we first need to understand what could  be recognized as a dynamical transition. Let say $determinism$ exists between two consecutive time points i.e.,  there exists a smooth mapping $\varphi$ such that 
\begin{equation} 
{\bf{x}}_{j+1}=\varphi ({\bf{x}}_{j}).  
 \label{eq:eq4}
\end{equation}In the case of a dynamical transition, this determinism breaks down and then we must not have any such $\varphi$. If we fix $\tau=1$ and if there is no dynamical transition between $j$ and $j+1$. Then we expect for a finite time series $S_{j|j+1}$ to fluctuate close to a constant value specific to the mapping $\varphi$. This particular feature of  $S_{j|j+1}$ will be explained in some detail in the next section (also see Fig.~\ref{fig:dia1}, it gives a schematic representation of the above introduced measures and concepts).  If a dynamical transition occurs between $j$ and $j+1$, then this leads to substantially large and sharp fluctuations in $S_{j|j+1}$. Which in turn could be quantified by the variance of $S_{j|j+1}$ over 
a window of $n$ points and given as 

\begin{equation}\label{eq:eq6}
  \sigma_{S}^{2} = \left<  (S_{j|j+1} - \mu_{S})^2  \right>, 
\end{equation}
where $\mu_{S} = \left< S_{j|j+1} \right>$, and $\left< \cdot
\right>$ denotes an average over $n$ points. We call $\sigma_S$ as the \emph{fluctuation of similarity}. Our numerical experimentation with a variety of nonlinear models with different kinds of transition has shown that $\sigma_{S}$ is a robust measure to identify distinct dynamical regimes and corresponding transitions. It shows even  more subtle transitions comparing to the standard measure of Lyapunov  exponent, and its potential has been demonstrated using chaotic transitions in the logistic  map \cite{nmalik_epl1}. Here we will be analysing several other nonlinear models by using this measure.   

In the next section we will attempt to establish the relationship between $\sigma_{S}$  and the dimension of the attractor as a first order approximation. We will also be providing some numerical results to support our analytical arguments. This  will be followed up with a discussion on the relationship between the above introduced measures and the Lyapunov  spectrum of the system.  

\section{Relationship with attractor dimensions}  
\begin{figure*}
 \centering
  \includegraphics [width=2.0\columnwidth]{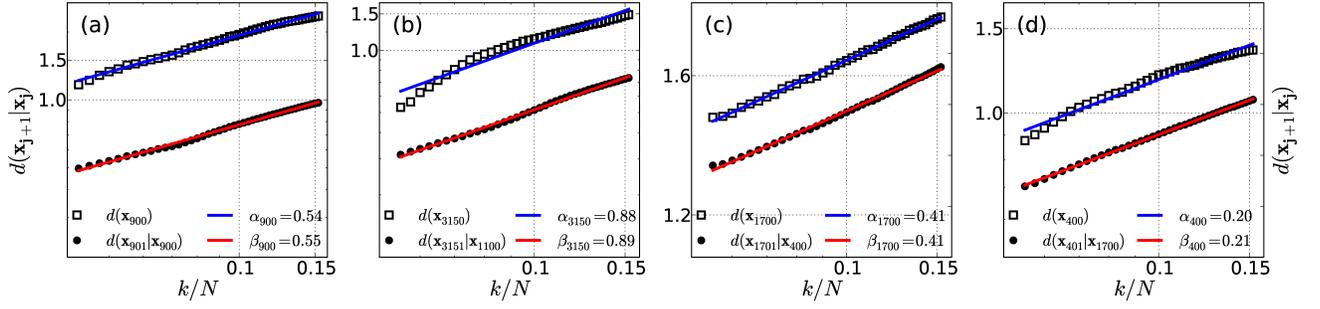}
  \caption{\small {(Color online) (a) and (b) Scaling laws for R\"ossler system for two
      different randomly chosen time points taken over a short time
      series of length $N=4500$. See Eq. (\ref{eq:eq1}) for red lines and  Eq. (\ref{eq:eq2}) for blue lines. Note that $\frac{\beta_j}{\alpha_j}  \to 1$ as we have not inserted dynamical transitions into the model. The embedding parameters used were $m=10$ and $L=15$.  (c) and (d) Scaling laws in the logistic  map for two different
      randomly chosen time points over a short time series of length
      $N=4500$. See Eq. (\ref{eq:eq1}) for red lines and  Eq. (\ref{eq:eq2}) for blue lines.  Note that again  $\frac{\beta_j}{\alpha_j}  \to 1$ as there are no dynamical transitions. The embedding parameters used were $m=3$ and $L=2$. Also, observe the fluctuations in the values of exponents $\alpha$ and $\beta$, caused by the shortness of series and numerical inaccuracies. }} 
  \label{fig:ross1}
\end{figure*}
The method presented above relies on  comparing dynamical similarity of two consecutive
time points in the embedded space, namely we only need to
calculate $S_{j | j+1}$ for intended application. To get detailed
insights into the properties of $S_{j | j+1}$, we make use of scaling
laws that exist for $d({\bf{x}}_{j})$ (the mean
distance of point ${\bf{x}}_{j}$ to its $k$ nearest neighbours) and
$d({\bf{x}}_{j+1} | {\bf{x}}_{j})$ (the mean distance of
${\bf{x}}_{j+1}$ to the $k$ nearest neighbours defined by the neighbourhood of ${\bf{x}}_{j}$), in
case model/map ~(\ref{eq:eq4}) is true \cite{arnhold1}. Suppose that a vector
${\bf{x}}_{j}$ in phase space has $k$ nearest neighbours then for
$k \ll N$, we will have the following scaling law (for further extensive
analytical details cf. \cite{dim-prl1,arnhold1,pettis1,parker1,dim-df1,dim-df2}): \begin{equation}
\frac{d({\bf{x}}_{j})}{\overline{d(}{\bf{x}}_{j})} = a_j(k/N)^{\alpha _j}, 
\label{eq:eq1}
\end{equation}
where $N$ is the length of the time series,  
 $a_j$ is a scaling coefficient and  $\overline{d(}{\bf{x}}_{j}) $ is the mean density of the whole
point cloud around $\bf{x}_j$, i.e., $\overline{d(}{\bf{x}}_{j}) =
\frac{1}{N} \sum_{k=1}^{N} \|{\bf{x}}_{j} - {\bf{x}}_{k} \|$. For $N \to
\infty$, $\alpha_j=D_F$, where $1/D_F$ is the effective dimension of the attractor. $D_F$ was first introduced in \cite{dim-prl1} and it has been conjectured in \cite{dim-df1,dim-df2} that $D_F$ is related to the $q$th order Renyi dimension $D_q$ by the following implicit relationship 
\begin{equation}
1=(q-1)D_q~~~:~~~\frac{1}{D_F}=D_q.
\label{eq:eq1d}
\end{equation}
For a stochastic time series $1/D_F=m$ where $m$ is the embedding dimension.
 
As the conditional distance between
${\bf{x}}_{j}$ and ${\bf{x}}_{j+1}$, namely $d( {\bf{x}}_{j+1} | {\bf{x}}_{j})$
also has a similar geometric formulation as the distance $d({\bf{x}}_{j})$,
hence conditional distance $d( {\bf{x}}_{j+1} | {\bf{x}}_{j})$ also scales with the ratio
$k/N$, and we can write 
\begin{equation}
\frac{d( {\bf{x}}_{j+1} | {\bf{x}}_{j})}{\overline{d(}{\bf{x}}_{j+1}| {\bf{x}}_{j})}=b_j
(k/N)^{\beta _j}, 
\label{eq:eq2}
\end{equation}
where $b_j$ is a scaling coefficient. In Fig.~\ref{fig:ross1}(a-d) we have numerically  demonstrated the above stated scaling laws for $d({\bf{x}}_{j})$ and $d( {\bf{x}}_{j+1} | {\bf{x}}_{j})$, by employing two different nonlinear systems. The first one is the R\"ossler system described by the following set of equations 
\begin{equation}
 \dot{x}  = -y - z;  
 \dot{y}  = x + ay;    
 \dot{z}  = 0.3x - 4.5z+xz    
\label{eq:eqross}
\end{equation}
where parameter $a=0.39$ corresponds to screw
type chaos (see Fig. ~\ref{fig:ross1}(a-b) for scaling behaviour). The second one is the logistic  map described by  
\begin{equation}
  x_{i+1}\! = \! 4 x_{i}(1-x_{i}) 
\end{equation} 
and corresponding scaling behaviour is plotted in Fig. ~\ref{fig:ross1}(c-d). Generally such scaling laws require  extremely large amount of data points \cite{small-bk,holger,pgrass_d1,pgrass_d2,pgrass_d3,gao1999,yong_power1},  but here we have attempted to obtain them using smaller amount of data points, namely with a time series of length $N=4500$. In Fig.~\ref{fig:ross1} we can clearly observe that scaling laws introduced in Eq.~(\ref{eq:eq1}) and Eq.~(\ref{eq:eq2}) hold even for short time series, though there are fluctuations in the values of the exponents. Therefore, in case of short time series we assume that $\alpha_j=D_F+\delta_j$, where $\delta_j$ are fluctuations due to the shortness of the time series and numerical
errors. Our attempt here is to provide the relationship between $D_F$ and $\sigma_S$ under the constraint that we are only considering short time series, i.e., for finite value of $N$.

Since the definition of the similarity between two consecutive time points is
${S_{j | j+1}=\frac{d({\bf{x}}_{j})}{d( {\bf{x}}_{j+1} | {\bf{x}}_{j})}}$, we
write the scaling law for similarity by taking into account Eqs.
(\ref{eq:eq1},~\ref{eq:eq2}) in the following form
\begin{equation}
 S_{j | j+1} = \overline{S}_{j | j+1}A_j (k/N)^{\gamma _j}, 
 \label{eq:eq3}
\end{equation} 
where $\gamma_j=\alpha_j-\beta_j$, ${\overline{S}_{j | j+1}=\frac{ \overline{d(}{\bf{x}}_j)}{\overline{d(}{\bf{x}}_{j+1}|
 {\bf{x}}_{j})} }$ and ${A_j=\frac{a_j}{b_j} }$. The dynamical similarities between
two consecutive time points ${\bf{x}}_{j}$ and ${\bf{x}}_{j+1}$ will
be determined by the relationship between the exponents $\alpha_j$ and
$\beta_j$. If no abrupt transition has occurred at the time
point $j$ then determinism should exist between time points
$j$ and $j+1$, i.e., a mapping of the kind $\varphi$ exists and Eq.
(\ref{eq:eq4}) holds. Then we expect $\gamma_j \approx 0$, i.e., $\beta_j
\approx \alpha_j$ if $N$ is infinity. In other words
$d({\bf{x}}_{j})$ and $d( {\bf{x}}_{j+1} | {\bf{x}}_{j})$ are expected to scale by the
same exponent. We provide an intuitive explanation of this in the sketch in
Fig.~\ref{fig:dia1}. Locally at ${\bf{x}}_{j}$ the mapping $\varphi$ can be approximated to be a linear transformation, which means that the neighbourhood $U({\bf{x}}_{j})$ will be deformed into an ellipsoid due to the application of mapping $\varphi$. Any expansion in the ball $U({\bf{x}}_{j})$ by inclusion of more points will lead to rescaling of the size of $\varphi (U({\bf{x}}_{j})$ by stretching or contraction in different directions. Hence, we will observe scaling of radii of these balls with the same exponent. We also expect in Eq. (\ref{eq:eq3}) that 
$\overline{S}_{j | j+1}A_j \to \text{const}.$ if $N \to \infty$. 
Hence, we could also say that $S_{j | j+1} \to \text{const}.$ for $N \to
\infty$. For rigorous mathematical expression for
${\overline{d(}{\bf{x}}_j)} / {\overline{d(}{\bf{x}}_{j+1}| {\bf{x}}_{j})}$ c.f.
\cite{pettis1,dim-df1,dim-df2,parker1}. The important point to note is that all
these scalings are only asymptotically valid. In the practical case of time series
of finite length, we observe
fluctuating deviations of the exponents of the scaling, similar to what we have observed in our numerical examples in Fig. ~\ref{fig:ross1}. Next we will
attempt to study the influence of these fluctuations on our method and
find an approximate expression for $\sigma_S$, the measure used
for identifying transitions. For convenience, lets define a variable $r_j$ such that  
${r_j^{\gamma_j}= \overline{S}_{j | j+1}A_j}$. Then Eq.~(\ref{eq:eq3}) can be written as   
\begin{equation} 
 S_{j | j+1} = (r_jk/N)^{\gamma _j}.
\label{eq:eq4b}
\end{equation}  In the considered examples we did not introduce any dynamical transitions hence determinism holds between two consecutive time points, and we do observe $\beta_j
\approx \alpha_j$ in Figs.~\ref{fig:ross1}. For both cases of R\"ossler system and logistic  map we obtained  $\frac{\beta_j}{\alpha_j} \to 1$ as expected.  Further we can write
  $\gamma_j = (1-\frac{\beta_j}{\alpha_j}) \alpha_j$, defining $\Delta_j=1-\frac{\beta_j}{\alpha_j}$. Therefore, for
  the case when determinism holds then $\Delta_j \to 0$ for $N$ tending to infinity.  As 
  mentioned above that $\alpha_j = D_F + \delta_j$, therefore we can write $\gamma_j
  =\Delta_j D_F + \Delta_j \delta_j$. Substituting this form of $\gamma_j$ in Eq. (\ref
  {eq:eq4b}) we get 
   \begin{equation}
 S_{j | j+1} =(r_jk/N)^{\Delta_j D_F+\Delta_j \delta_j}.
 \label{eq:eq5}
\end{equation}
As $\Delta_j$ and $\delta_j$ are small terms, we can neglect their
product. Then we log transform  Eq.~(\ref {eq:eq5}) to finally yield
\begin{equation}
S_{j | j+1} = \exp{(D_F \ln(r_jk/N)\Delta_j)}.
 \label{eq:eq6}
\end{equation}

Expanding the right hand side of Eq. (\ref {eq:eq6}) in terms of
exponential series and neglecting the higher order terms in $\Delta$, 
 we get
 \begin{equation}
  S_{j | j+1}   = 1+D_F\ln(r_jk/N)\Delta_j.  \nonumber 
\end{equation}
Writing $\Delta'_j =\Delta_j (\frac{\ln r_j}{ \ln (k/N)}+1) $, the above
expression becomes
   \begin{equation}
  S_{j | j+1}   = 1+D_F\ln (k/N)  \Delta'_j.  \nonumber 
\end{equation}
Therefore the average of  $ S_{j | j+1}$ taken over a window of size $n$ is,  
\begin{equation}
\mu_S=\left< S_{j | j+1} \right>= 1+D_F\ln (k/N)  \left<\Delta'_j \right>. 
\nonumber
\end{equation} 
Finally, for the standard deviation we obtain the following expression,
 \begin{equation}
 \sigma_{S}^2={(D_F \ln (k/N))}^2\sigma_{\Delta'_j}^2. 
 \label{eq:eq9}
\end{equation} Equation~(\ref{eq:eq9}) shows the explicit dependence of $\sigma_S$
on the dimension $D_F$ of the attractor (commonly  $1/D_F$ is referred as the effective dimension of the attractor \cite{arnhold1}), since the term $\ln k/N$ is kept constant over the whole length of the time series. \emph{Consequently, changes in the
structure of the attractor will lead to changes in the value of $\sigma_S$}.

\begin{figure}
 \centering
  \includegraphics [width=\columnwidth]{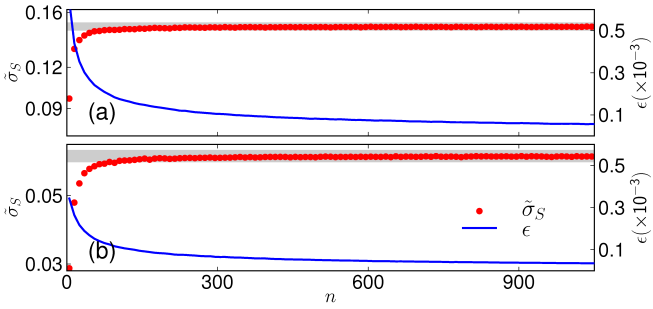}
  \caption{\small {(Color online) Convergence of  $\sigma_{S}$ . $\tilde{\sigma}_{S}$ is the median 
     over $10,000$ bootstrap realizations of $\sigma_{S}$ and $\epsilon$  gives the corresponding standard error in the estimation of the median over these realizations. $n$ is the window size over which $\sigma_{S}$ has been calculated. 
     (a) Logistic map (b) R\"ossler system (in both cases, the
    total length of time series is $N=4500$.) }} 
  \label{fig:conse}
\end{figure}

The law of large numbers must lead $\sigma_{\Delta'_j}^2$ to converge,  
as the two constituents of $\Delta'_j $, i.e. $ r_j $ and $ \Delta_j $ are themselves
expected to converge to constant values for large $N$. If the fluctuation term $\sigma_{\Delta'_j}^2$ converges for a large enough window size, then $\sigma_S$ will also converge to a value which is a multiple of the attractor dimension. Let us next put this argument to a numerical test, in order to answer the question whether an increase in the number of observations in calculation of $\sigma_S$ leads to convergence (Fig.~\ref{fig:conse}) ? 
In Fig.~\ref{fig:conse} we show this convergence of
$\sigma_S$.  In Fig.~\ref{fig:conse}(a) we consider
the logistic  map time series of length $N=4500$. We calculate $\sigma_S$ with increasing window size $n$ i.e., including increasing number of time points in the calculation of $\sigma_S$. 
The median of $\sigma_{S}$ i.e., $\tilde{\sigma}_{S}$ is calculated for window size $n$ 
taking $10,000$ realisation of $\sigma_{S}$ by boot strapping. The value of
$\tilde{\sigma}_{S}$ quickly converges as $n$ the window size is increased. The  standard error $\epsilon$ in calculation of $\sigma_{S}$ also show a continuous drop before saturating to small value around $0.0002$. A similar conclusion was reached
for the R\"ossler system  in Fig.~\ref{fig:conse}(b). This also supports
the usefulness of the windowing technique we have used in this work.

Our extensive numerical experimentation has demonstrated that $\sigma_S$ is extremely sensitive to changes in the dynamics. The reason for this seems to be that any dynamical transition will lead to the breakdown of determinism between consecutive time points, i.e., Eq. (\ref{eq:eq4}) will not be valid anymore. This simply means that $\gamma_j \nrightarrow 0$
(or $\frac{\beta_j}{\alpha_j} \nrightarrow 0 $), which in turn will produce
a  large fluctuation in the values of $S_{j | j+1}$.  These
fluctuations will be captured by $\sigma_S$. The statistically most
significant fluctuations indicate  dynamical transitions,
and could be identified by means of statistical significance tests as
described later in Sec.\ref{sec:stat-trans}. We will continue this discussion about
the analytical properties of $S_{j | j+1}$ and its average and variance
in the Section~\ref{sec:lypsec1}. Next we present a numerical example to demonstrate, that $\sigma_S$  is sensitive to changes in the dimension or complexity of the attractor. 
 
 \begin{figure}
 \centering
  \includegraphics [width=\columnwidth]{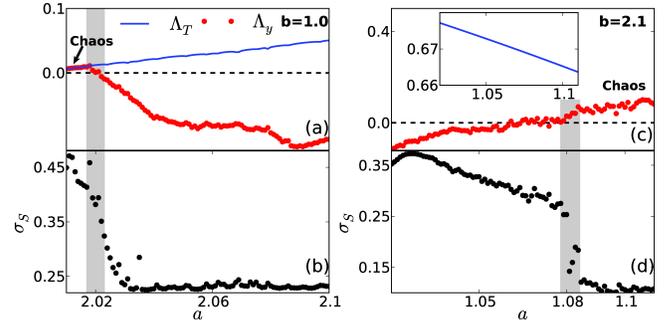}
    \caption{\small {(Color online) (a,b) Transition from Chaos to SNA  highlighted by gray vertical band and (c,d) transition from SNA to chaos. (a,c)  Largest transverse Lyapunov  exponent
$\Lambda_T$ and Largest Lyapunov  exponent $\Lambda_y$  
  (b) $\sigma_S$ shows sharp drop in its value as the SNA appears.
  similarly in (d) $\sigma_S$ show a sharp drop as SNA disappears into chaos. }}
  \label{fig:snat}
\end{figure}

The strange non-chaotic attractors (SNA) appear in various 
quasi-periodically driven dissipative dynamical systems
\cite{sna01_1,tolga,ying}. The transition between chaos and SNA are quite subtle and identifying them is a challenging numerical problem \cite{jojo1,jojo2}. Here, we apply the presented method for identifying dynamical transitions to and from SNA. We will attempt to
identify transitions in a coupled map of the form:
 \begin{eqnarray}
 x_{i+1}&=& (x_{i}+2\pi \omega)  \mod(2\pi),  \\  \nonumber 
 y_{i+1}&=& \frac{1}{2\pi} (a \cos(x_i)+b)\sin(2\pi y_i).
 \label{eq:eqsna}
 \end{eqnarray}
The two types of Lyapunov  exponents namely, the largest transverse Lyapunov  exponent
$\Lambda_T$ and the largest Lyapunov  exponent $\Lambda_y$ of the subsystem $y$, are
given by the following set of equations~\cite{tolga,ying}
\begin{eqnarray}
\Lambda_T&=&\lim_{n\to \infty} \frac{1}{n} \sum_{j=1}^{n} \ln | a\cos(x_j)+b|,  \\
\nonumber 
\Lambda_y&=&\lim_{n\to \infty} \frac{1}{n} \sum_{j=1}^{n} \ln |
a\cos(x_j)+b\cos(2\pi y_j)|.
\label{eq:eqsna01}
\end{eqnarray} 
It is known that in the case of $\Lambda_T > 0$
and $\Lambda_y < 0$ we have SNA while for $\Lambda_T >
0$ and $\Lambda_y > 0$ we have a chaotic regime. In Fig.~\ref{fig:snat}
(a,~b) the grey band represents the transition to SNA from chaos. 
This transition known to occur via on-off intermittency. Whereas the grey band in Fig.~\ref{fig:snat}
(c,~d) highlights the transition from SNA to chaos. 

We generate a short time series of length $N=4500$ at $100$ different
values of $a$ separated by $0.002$. Then we calculate $\sigma_S$ using
embedding parameters $m=5$ and $L=2$. In Fig.~\ref{fig:snat}  we have
plotted $\sigma_S$ with the $\Lambda_T$ and $\Lambda_y$. An abrupt change in the values of $\sigma_S$ would indicate a transition. Comparing  Fig.~\ref{fig:snat} (a,~b) we observe that as $\Lambda_y$  starts to decrease and becomes negative, the values of $\sigma_S$ show a simultaneous drop, signifying the dependence of $\sigma_S$ on the complexity or qualitative features of the dynamics. We observe lower values of $\sigma_S$ for SNA than for chaos. A similar change is observed if we reverse this transitions i.e., going from SNA to chaos (Fig.\ref{fig:snat} (c,~d)). As the values of $\Lambda_y$ increase to positive values there is again a sharp drop in the values of $\sigma_S$. This example demonstrates that $\sigma_S$ is able to capture even a subtle change in dynamics, like the ones that occur in transitions between the SNA and the chaos. In \cite{nmalik_epl1} we have shown that $\sigma_S$ can uncover all the transitions that are induced by the variation of the parameter in a logistic  map like period-chaos transitions, intermittency, chaos-chaos transitions, etc..

\section{Relationship with Lyapunov  spectrum} \label{sec:lypsec1}

Lyapunov  exponents $\lambda_i$ are the most extensively used measures for a
quantitative characterization of nonlinear dynamics \cite{ruelle_1,small-bk,holger,parker1,hab1}. Several dynamical invariants are conjectured in terms of them such as Lyapunov  dimension. However, a reliable numerical method to estimate $\lambda_i$ from short time series
remains to be a challenging problem~\cite{wolf1,parker1,jk001}, which we frequently encounter
in various real time systems. The main objective of this section is to
understand the new measure $S_{j | j+1}$, its mean $\mu_S$ and variance $\sigma_S$ in terms of
these well known dynamical measures of Lyapunov  exponents.

\begin{figure}
 \centering
 \includegraphics [width=\columnwidth]{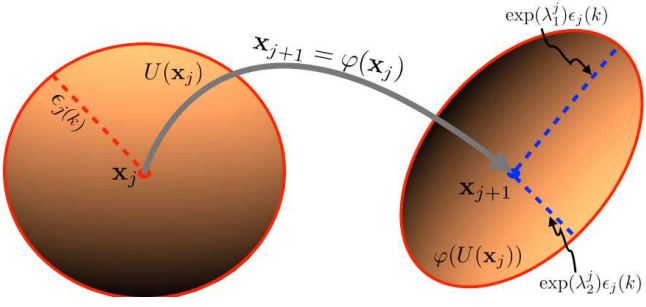}
 \caption{\small {(Color online) Evolution of the $\epsilon(k)$ neighbourhood of the
     time point ${\bf{x}}_{j}$ into an ellipsoid by the
     application of the smooth mapping $\varphi$ such that
     ${\bf{x}}_{j+1}=\varphi ({\bf{x}}_{j})$. The expansion or
     contraction in any direction $i$ is a multiple of
     $\exp(\lambda_i^j)$, where $\lambda_i^j$ are the eigenvalues of
     ${\bf{D}}\varphi$ at $j$. }}
    \label{fig:dia2}
    \end{figure}

Suppose that Eq.~(\ref{eq:eq4}) holds and 
$\lambda_1^j, \lambda_2^j,\dots,\lambda_m^j$ are the eigenvalues
of the Jacobian matrix ${\bf{D}}\varphi({\bf{x}}_j)$. 
Then the deformation of the infinitesimal 
$\epsilon_j(k)$ ball neighbourhood of ${\bf{x}}_j$ in any direction $i$
will be multiple of $\exp(\lambda_i^j)$  (see Fig.~\ref{fig:dia2}).
Defining ${\Lambda_i^j=\exp(\lambda_i^j)}$, where $\Lambda_i^j$ are
called the \emph{Lyapunov  numbers}. The \emph{local Lyapunov 
  exponents}, $\lambda_i$ are given by
\begin{equation}
\lambda_i(n)=  \frac{1}{n}\displaystyle\sum_{j=1}^{n} \lambda_i^j.  
\label{eq:eqhi0}
\end{equation} 
The \emph{global Lyapunov  exponent} $L_i$ corresponding to the direction
$i$ is the asymptotic value of $\lambda_i$
\begin{equation}
L_i= \lim_{n \to \infty} \frac{1}{n}\displaystyle\sum_{j=1}^{n} \lambda_i^j. 
\label{eq:eqhi1}
\end{equation} 
If the distance metric used for calculation of $d({\bf{x}}_{j})$ is the
Euclidean then a simple geometrical consideration yields
\begin{equation}
 d({\bf{x}}_{j+1} | {\bf{x}}_{j})= \left( 
  \frac{1}{m} \sum_{i=1}^{m}   {\Lambda_i^j}^2 ~
 \right)^{\frac{1}{2}} d({\bf{x}}_{j}),   \nonumber
\label{eq:eqhi2}
\end{equation}  
which directly leads to  
 \begin{equation}
  S_{j | j+1}   =  \left( \frac{1}{m} \sum_{i=1}^{m}   {\Lambda_i^j}^2 ~ \right)^{-\frac{1}{2}}.
 \label{eq:eqss1}
\end{equation}
Hence, $S_{j | j+1}$ measures the total deformation of the $\epsilon_j(k)$ ball
neighbourhood of point ${\bf{x}}_{j}$ when a mapping $\varphi$ is applied on it.

\begin{figure}
 \centering
  \includegraphics [width=\columnwidth]{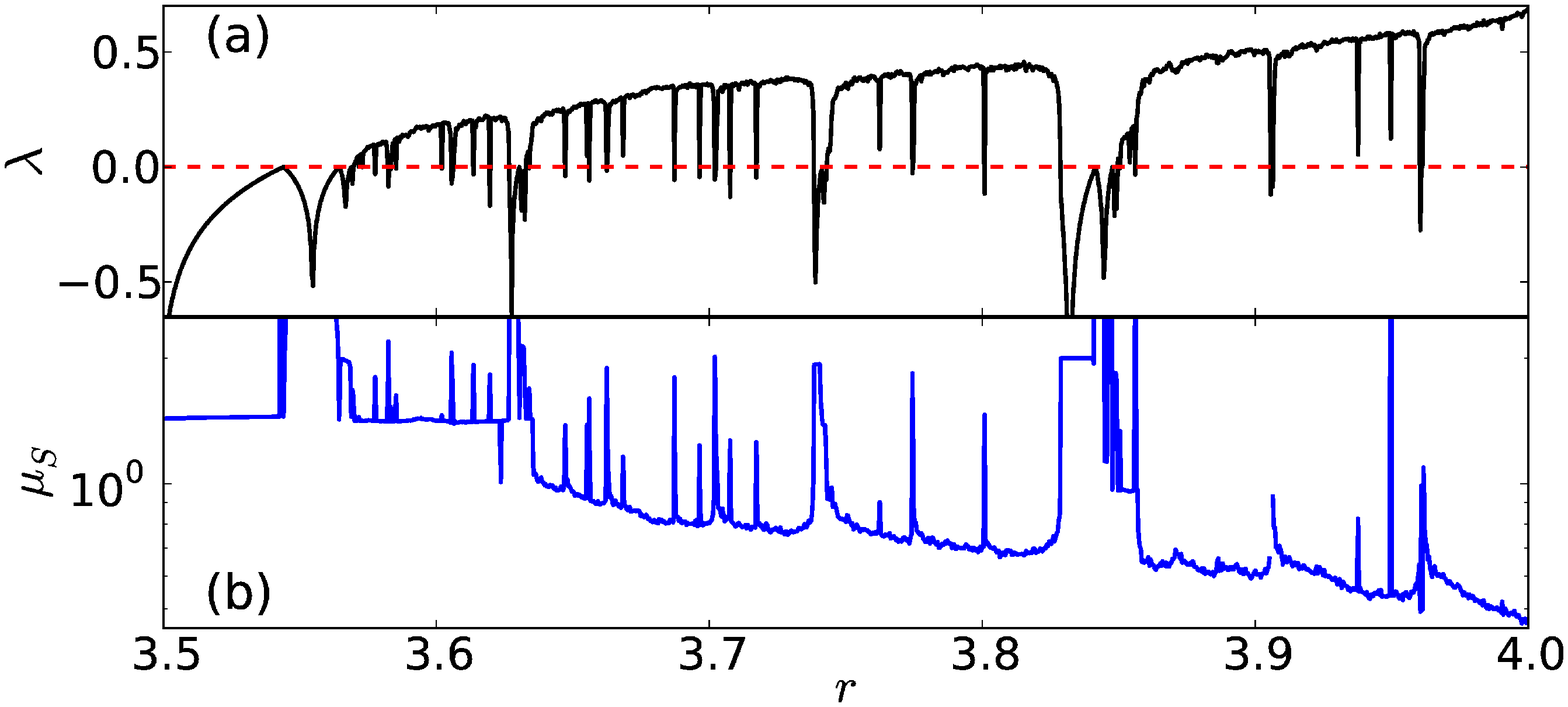}
  \caption{\small {(Color online) Correspondence between $\mu_S$ (blue
  line in (b)) and the Lyapunov  exponent $\lambda$ (black line in (a)) for the logistic map.
  The parameters used for this figure are the same as used for Fig.~\ref{fig:ross1}.}}
    \label{fig:flyp1}
\end{figure}

From Eq.~\ref{eq:eqss1}, we find that the average of $S_{j | j+1}$ taken over a window of size $n$ is,
\begin{equation}
\mu_s= \frac{1}{n}\displaystyle\sum_{j=1}^{n} \left[ \left( 
  \frac{1}{m}  \sum_{i=1}^{m}  {\Lambda_i^j}^2 ~
 \right)^{\frac{1}{2}} \right].
\label{eq:eqhi3}
\end{equation}
Comparing, Eq.~(\ref{eq:eqhi3}) with Eq.~(\ref{eq:eqhi0}) and Eq. ~(\ref{eq:eqhi1})  we can 
conclude that  $\mu_S$  will be structurally the same as the sum of the  local
Lyapunov  exponents, while $\mu_S$ over large $n$ will be structurally the same
as the sum of the global Lyapunov  exponents. This is shown numerically for
the Logistic map in Fig.~\ref{fig:flyp1}.  In case of a chaotic system we always have a direction  $i$ such that the 
 $\lambda_i^j > 0$ i.e., $\Lambda_i^j >1$,  representing the expansion in the direction $i$. In other directions we will either have contraction, 
 $\lambda_i^j<0$, i.e., $\Lambda_i^j<1$ or  $\lambda_i^j=0$, i.e., $\Lambda_i^j=1$. Therefore, in a chaotic system with few degrees of freedom  the most dominant contribution to $S_{j | j+1}$ in Eq.~(\ref{eq:eqss1}) comes from the largest
positive eigenvalue corresponding to the expansion. Hence, for such a systems $\mu_S$ will be structurally similar to the Lyapunov exponent. In Fig.~\ref{fig:flyp1} we observe a structural correspondence between  $\mu_S$ and the Lyapunov exponent of the logistic map in form of  an anti-phase relationship. We know the sum of the largest Lyapunov  exponents is proven for certain systems to be related with dynamical invariants such as Lyapunov dimension, topological entropy, and information dimension (due to Kaplan-Yorke conjecture)  ~\cite{ruelle_1,dim1_1,dim2_2,ottbook}.  Therefore, we may think  that $\mu_S$ could also be used in quantifying dynamics but our numerical analysis 
has shown that $\mu_S$ is not well suited for detecting dynamical transitions in the series because it is less sensitive in quantifying and capturing large fluctuations in $S_{j | j+1}$. It will be better to use $\sigma_S$ for this
purpose, also $\sigma_S$ is better suited to quantify variation in parameters of
a dynamical system and corresponding changes in the dynamics, because any
variation in the parameters of a dynamical system will lead to variation in the
Lyapunov  spectrum too.  Transitions lead to large deformations of the $\epsilon_j(k)$ neighbourhood, leading  to large fluctuations in the magnitude of $S_{j | j+1}$, which are then captured by $\sigma_S$. A  point to note is that the global
Lyapunov  exponents  $L_{i}$ are not defined for a time series with a transition, due to the constraints imposed by ergodicity.

In order to get further insights into the properties of measure $\sigma_S$ we
take a look at the distribution of $S$ (dropping subscript for
simplicity) over window size $n$ of vectors, $P(S,n)$.  
$\lambda_i^j$ are in a sense random numbers for chaotic systems. The
expression of $S$, Eq.~(\ref{eq:eqss1}) consists of a summation over
$\Lambda_i^j=\exp(\lambda_i^j)$, therefore, the central limit
theorem implies that $S$ follows a Gaussian distribution at least
asymptotically. Following~\cite{ottbook}, we find that
asymptotically $P(S,n)$ has the following general analytical form,
\begin{equation}
   P(S,n) \sim \displaystyle\frac{1}{  \displaystyle \sqrt{2\pi n \Phi^{n}(S) } 
   }    \exp{(-n \Phi (S))},
\label{eq:disss1}
\end{equation}        
where $\Phi(S)$ is a convex quadratic function with minimum zero,
occurring at $S=\mu_S$ i.e., $\Phi(\mu_S)=0$ also, $\Phi'(\mu_S)=0$,
$\Phi''(\mu_S)>0$. Expanding $\Phi(S)$ around $\mu_S$ and neglecting
higher order terms, we write Eq.~(\ref{eq:disss1}) as,
\begin{equation}
   P(S,n) \sim \displaystyle\frac{1}{  \displaystyle \sqrt{2\pi n \Phi''(S) }}  
   \exp{(-n\Phi''(S) \frac{(S-\mu_S)^2}{2})},
  \label{eq:disss2}
\end{equation} 
which gives a familiar looking form of a Gaussian distribution. Then
\begin{equation}
\sigma_S=(n\Phi''(S))^{-1/2}. 
\end{equation} A similar expression for distribution
and variance could also be written for the local Lyapunov  exponents
\cite{ottbook}, where $\Phi( \lambda_i)$ is known as the spectrum of the 
local Lyapunov  exponents and can be used for characterising the
dynamics of the system~\cite{grass11,sep1,awd1}. So as an analogy we propose
that $\Phi(S)$ can also be used to characterise the dynamics. The distribution
of $S$ for different types of dynamics may follow a Gaussian distribution of the
type $P(S,n)$ asymptotically but each type of dynamics must correspond to unique
$\mu_S$ and $\sigma_S$. This is because of the fact that each type of dynamics
has a unique $\varphi$ (see Eq.~(\ref{eq:eq4})) and hence unique eigenvalues and the
corresponding deformations and values of $S$ should also be unique. In
future research we intend to develop a method based on estimation
of $\Phi(S)$ to classify distinct dynamics.

\section{Dynamical transition induced by co-evolving parameters}  \label{sec:stat-trans}

In the numerical example above values of $\sigma_S$ could distinguish between two distinct chaotic regimes and SNA, demonstrating that $\sigma_S$ can be used to distinguish different types of dynamics.  A similar conclusion about the capability of $\sigma_S$ in identifying distinct dynamics could be made from an example of logistic  map  presented in \cite{nmalik_epl1}, where $\sigma_S$ was used to uncover all the transitions that are induced by the variation of the parameter in a logistic  map like period-chaos transitions, intermittency, chaos-chaos transitions, etc.. One important point is that in these examples we had one whole time series at each value of the parameter. In many realistic systems we do not have the luxury of a whole time series being available at a single value of the control parameter. Rather, the most common real situation is when control parameter also co-evolves with the dynamics \cite{scheffer,dbif_3,dbif_1,dbif_2}. We only have very few points available at a particular value of the control parameter. For example, in palaeoclimate  we have few observation of a climatic variable via proxies,while parameters which drive climate like solar insolation co-evolve with these climatic variable at time leading to transitions in the dynamics \cite{nmalik_epl1,lenton,ashwin,mos1,dbif_2}. Another example of this situation is observed in social dynamics, where we have very few observation of social indices while the parameters driving social dynamics, like the economic and political situations, coevolve with it \cite{scheffer,ec1,econ_1,econ_2}. A further example of this situation is in neuroscience, where event-related potentials (ERP) measured by EEG  show several distinct dynamical behaviours as a  response to changing stimuli \cite{jkbook1,epr1}.  One possible conceptual model for such transitions could be: 
\begin{equation}
\dot{\bf{y}}=f(\bf{y},\zeta(t)),
\label{eq:dbif}
\end{equation} where $\bf{y}$ is a set of variables of a dynamical system, with $\zeta(t)$ being a parameter evolving with time. Rate of change of $\zeta(t)$, or passing of $\zeta(t)$ through the bifurcation point of the system can lead to a variety of qualitative changes in the dynamics of the system, including the more subtle one of shifting of the system from regime of one complex chaotic dynamics to other chaotic dynamics of higher or lower complexity\cite{dbif_1,dbif_2,dbif_3,ashwin,holger}. In the numerical examples following this section, we let the parameter simultaneously evolve with variables of the system. This would provide us more realistic model examples to test our method for its practical usefulness.

We have also introduced a statistical test for assisting a more automatized identification of dynamical transitions in \cite{nmalik_epl1}. For convenience we describe it here again : We have used the temporal evolution of $\sigma_S$ to identify the changes in dynamics. To test the relative statistical significance of two values of $\sigma_S$ to belong to distinct or same dynamics, we use a bootstrapping procedure, where we randomly draw $n$ values with replacement from the series of $S_{j|j+1}$, where $n$ is the window size used in calculation of $\sigma_S$. Repeating this procedure several thousand times we generate an ensemble of values of $\sigma_S$. Then we interpret  0.05 and 0.95 percent quantiles of this ensemble as the $90\%$ confidence bounds.  The values of $\sigma_S$ outside this bound are less probable to occur. Hence, we can classify these points as belonging to dynamics of two distinct complexity with $90\%$ confidence. The time band over which the crossover between the two levels occurs contains the point of dynamical transition. The points with lower values of $\sigma_S$ may be regarded as belonging to dynamical regimes which are relatively more stable and lower in dynamical complexity.

\subsection{Identifying drift in the dynamics (nonstationarity)}  

\begin{figure}
 \centering
  \includegraphics [width=\columnwidth]{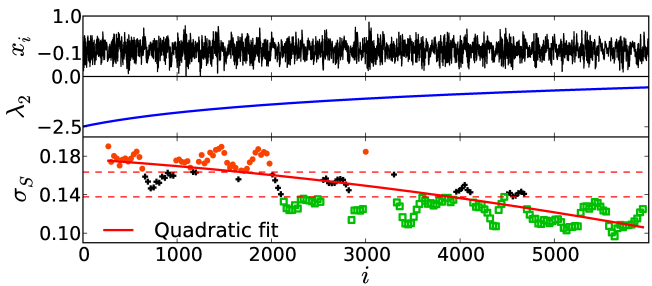}
  \caption{\small { (Color online) Identifying drift in the dynamics of the Baker's map
      with continuously changing parameter $\beta$: According to
      Eq.~(\ref{lambda1_baker}), $\lambda_1$ is constant; $\lambda_2$ is equal
      to $\ln \beta$, so it varies continuously in a nonlinear way. We     
      observe $\sigma_S$ changing from significantly higher values to
      significantly lower values as the time progress. A quadratic fit describes
      this evolution (red curve), indicating the nonlinear change of parameter
      of the system.}}
  \label{fig:bak}
\end{figure}

One of the challenging problems could be identifying a continuous
drift in the dynamics of a time series. For this purpose we use the
generalized Baker's map~\cite{farmer} and generated a time series
following the same procedure as described in \cite{schreiber1997}.
 \begin{eqnarray} 
 \mbox{if} ~ v_i  \leq \alpha: ~ u_{i+1} &=& \beta u_i  , ~~  v_{i+1}=v_i / \alpha,  \nonumber  \\
  \mbox{if}  ~ v_i > \alpha:   ~  u_{i+1} &=& 0.5+\beta u_i ,  ~~ v_{i+1}= \frac{(v_i-\alpha)}{(1-\alpha)}.
\end{eqnarray}
The Lyapunov  exponents for the above set of equations are 
\begin{eqnarray} \label{lambda1_baker}
\lambda_1&=& \alpha \ln \frac{1}{\alpha} +(1-\alpha) \ln \frac{1}{1-\alpha}, \\  
\lambda_2&=&\ln \beta.
 \end{eqnarray}

Now we introduce  a drift in the parameter $\beta$ as done in \cite{schreiber1997},
namely, by generating a time series of length $15,000$ by varying $\beta$ in
each iteration by $\beta = i/15,000$ and fixing the value of $\alpha = 0.4$. This
creates a nonstationary time series with drift in dynamics, while the maximal
Lyapunov  exponent $\lambda_1$ is constant $\alpha = 0.4$
(Eq.~\ref{lambda1_baker}). The trend from the time series is removed by taking
\begin{equation}
x_i = \frac{w_i- \left<w\right>_k}{ \sqrt{\left<(w_i- \left<w\right>_k )^2
\right>_k}},  \nonumber
\end{equation}
where $w_i = u_i + v_i$ and we took $k$ = 50. We consider only a short
section of the time series by taking points from $i = 1000$ to $i =
7000$, which means we have considered only $6000$ time points. However, in the
original work of~\cite{schreiber1997}, $40,000$ data points were used. We use the 
embedding parameters $m=5$ and $L=2$. The result is shown in
Fig.~\ref{fig:bak}, where we observe that values of $\sigma_S$ go from
significantly higher values to significantly lower ones, which is
representative of the dynamical drift that has taken over the time points.

\subsection{Transition between transient chaos and Lorenz's attractor}  
 
\begin{figure}
 \centering
  \includegraphics [width=\columnwidth]{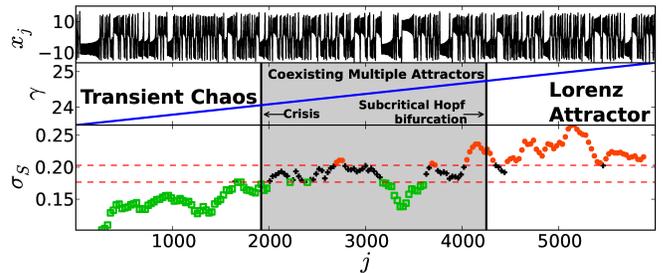}
  \caption{\small {(Color online) Formation of Lorenz's attractor, when parameter $\gamma$ is varied from $23.5$ to $25.25$ (see Eq.~(\ref{eq:eq-lr})). When $\gamma <24.06$, transient chaos exist whereas $24.06 < \gamma < 24.74$ is a \emph{transition zone} where multiple attractors coexist and for $\gamma > 24.746$ only Lorenz's attractor exists. The $\sigma_S$ distinguishes between these three regions, for $\gamma <24.06$ only green open squares points exist indicating low complexity dynamics,  for $\gamma >24.74$ only orange dots exist indicating higher complexity dynamics. In between these two regions we have a state where multiple attractors coexist, this is indicated by  jumps in the values of $\sigma_S$ below and above the significance bands (red dotted horizontal lines). The thick black vertical lines are drawn at the times points when $\gamma \approx 24.06$  (crisis) and  $\gamma \approx 24.74$ (subcritical Hopf bifurcation)}}
  \label{fig:lrfig}
\end{figure}

Another example somewhat similar to the above one, but in case of a time continuous system, is the formation of Lorenz's attractor from transient chaos :   
\begin{eqnarray} 
\dot{x} & = & a (y - x), \nonumber \\ 
\dot{y} & = & x (\gamma - z) - y,   \\
\dot{z} & = & x y - b z. \nonumber 
\label{eq:eq-lr}
\end{eqnarray} The chaotic Lorenz's attractor exists for $\gamma >24.74$, whereas transient chaos exists for $\gamma <24.06$. For the transition zone $24.06 < \gamma < 24.74$, there is a coexistence of three attractors:  two being steady states and one a chaotic attractor \cite{sparrow}. The transient chaos disappears due to a crisis at $\gamma \approx 24.06$ and Lorenz's attractor emerges as the only possible stable attractor due to a subcritical Hopf bifurcation at  $\gamma \approx 24.74$. We generate a time series of the $x$ variable by solving Eq.~(\ref{eq:eq-lr}), using a Runge-Kutta fourth order procedure at time step resolution of $10^{-3}$, while sampling a point after $10^3$ time steps. We have sampled $6,000$ time points and varied $\gamma$ linearly between $23.5$ to $25.25$. So, we can substitute $\gamma=\gamma_\circ+\Delta T t$ in Eq.~(\ref{eq:eq-lr}), where $\gamma_\circ=23.5$ and $\Delta T$ is a small increment of the order of $10^{-7}$. This variation leads the system to pass through the transient chaos to a transition zone (crisis and subcritical Hopf transitions)  to the formation of Lorenz's attractor (Fig.~\ref{fig:lrfig}).   

To calculate $\sigma_{S}$ we have used $m=10$, $L=10$ and a window size of $300$ with $90\%$ overlap. A detailed explanation for our choice of rather higher values of embedding parameters  is provided in Sec.~\ref{sec:stat-empar}. The calculated values of $\sigma_{S}$ are shown in the lowest panel of Fig.~\ref{fig:lrfig}, with color of markers standing the same as in previous example. We observe lower values of $\sigma_{S}$ (green open squares) below the confidence bound for transient chaos and higher values (orange dots) above the confidence bound for Lorenz's attractor, hence distinguishing both dynamical regimes in this time series. The transition zone (grey shaded region) not only contains multiple transition but also multiple attractors which is also reflected in the values of $\sigma_{S}$, while it jumps between green open squares and orange dots few times. Due to the fact that the formation of an attractor is temporally delayed \cite{delay-bif}, and we also loose few initial points due to windowing and embedding, the transitions are usually rightward shifted.

 \subsection{Tolerance of the measure against observational noise} \label{sec:noise-trans}

\begin{figure}
 \centering
  \includegraphics [width=\columnwidth]{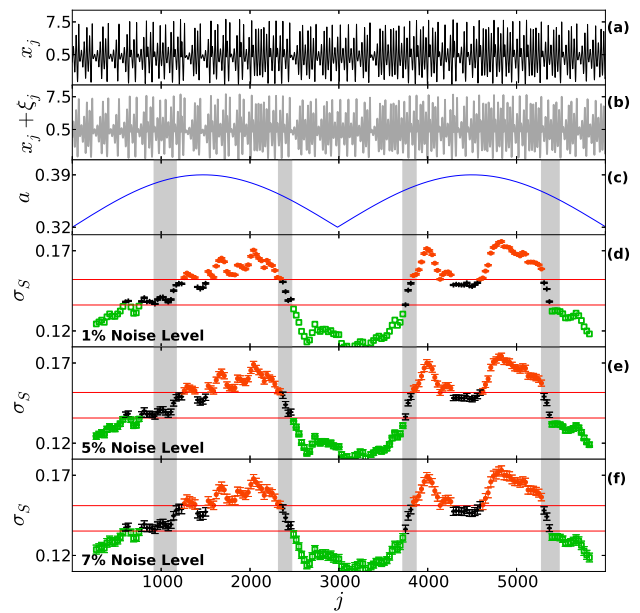}
  \caption{\small {(Color online) Test the effect of presence of observational noise on the method:
      (a) $x_j$ are values of $x$ variable of Eq.\ref{eq:eqross} sampled at time step $j$, (b) shows the effect of adding white noise to $x_j$. (c) shows the variation of control parameter $a$, note the crossing of four grey bands by the parameter $a$. These grey band represent the four dynamical transition that take place when $a$ crosses the value of $0.39$. (d-f) shows the variation of $\sigma_S$ with three different levels of noise added to the signal. For $1\%$ we see the smallest error bars and all the four transitions are clearly visible, i.e., crossing of significane band by values of $\sigma_S$ right between the grey bands. In higher noise levels the transitions are still intact but with increasing error bars.}}
  \label{fig:nos}
\end{figure}

To test the influence of observational noise on the above introduced measure, we consider the example of the R\"{o}ssler model Eq.~(\ref{eq:eqross}). In this system two topologically distinct attractors exist, namely spiral type chaos for   $0.32\leq a < 0.39$ and screw type chaos for $a\geq 0.39$ \cite{ros1,ros2}. The transition behaviour occurs via the formation of a  homoclinic orbit at $a \approx 0.39$ \cite{ros1}. We generate a test time series for our method by varying the control parameter $a$ and by defining its temporal evolution as $a(t) = 0.32 + 0.07| \sin( \pi \Delta t)|$ at every six hundredth integration step. $\Delta t$ is the step size for the fourth order Runge-Kutta integrator ($\Delta t = 0.001$). Then we sample $6,000$ points of the $x$-component at the rate of  $200\Delta t$. This leads $a$ to cross the  transition point four times (see Fig.~\ref{fig:nos} (c)). This example was also discussed in \cite{nmalik_epl1}. Here we discuss it in the context of presence of observational noise in this section and missing values in the next section.

To add white noise into the time series we generate normally distributed random variable $\xi$ with its mean $\langle \xi \rangle=0$ and its standard deviation $\sigma(\xi)=\eta\sigma(x)$, where $\sigma(x)$ is the standard deviation of the whole time series. Then we simply add a $\xi$ to each value of $x$ in the time series (see Fig.~\ref{fig:nos} (b)). We can vary the strength of noise by varying $\eta$, for instance when $\eta=0.01$ we have $1\%$ noise level or $20dB$ noise in the signal. We test the tolerance of the measure against three different noise levels here, viz. $1\%$, $5\%$, and $7\%$ (see Fig.~\ref{fig:nos} (d-f)). The error bars on the values of $\sigma_S$ are obtained by generating $1,000$ different realizations of noise at each level. We have replaced the significance levels from dotted red lines to solid red lines, as these are the mean of the significance levels for all the different realizations of the noise. In Fig.~\ref{fig:nos}(d-f) we observe that all the transitions seems to remain intact for all the different levels of noise. This is a clear indication that this method is robust against nominal levels of noise. We have also attempted the above numerical experiment with some other models, and the results of those experiments also demonstrate similar robustness of this method against nominal levels of noise. The embedding parameters used for every level of noise are exactly the same,  we had set $m = 10$ and $L = 15$ and window size of $300$ with $90\%$ overlap. These parameters are also the same as used in \cite{nmalik_epl1}, while discussing this example in noise free case.

\subsection{Strategy for treatment of missing values} \label{sec:missing-trans}

\begin{figure}
 \centering
  \includegraphics [width=\columnwidth]{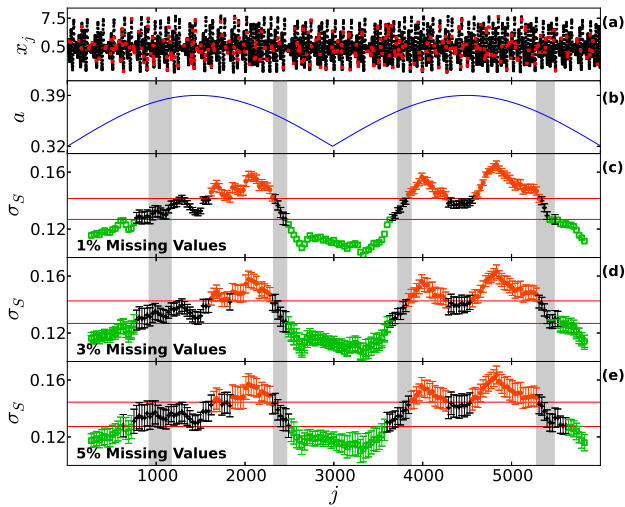}
  \caption{\small {(Color online) Test the effect of missing values on the method:
     (a) Red points represent the time points missing from the dynamics
      (black points) of the R\"ossler system at the level of $5 \%$ missing values. 
     The  transition highlighted by grey band are introduced by changing the parameter $a$ in Eq.
      \ref{eq:eqross} as described above in Sec. \ref{sec:noise-trans}. 
      (c-e) show the variation of  $\sigma_S$ at different levels of missing values, in all the three we observe the transitions to remain intact. A important point note is that even on using Chebyshev distance, there is no structural change in evolution of $\sigma_S$  compared to previous example (See Fig.~\ref{fig:nos}). }}
  \label{fig:miss}
\end{figure}

Apart from shortness of the data, another central problem which surrounds data analysis is irregular sampling or missing values \cite{irr1,irr2,irr3}. This is a common problem in fields such as astronomy, medical, earth, and social sciences \cite{irr1,irr2,irr3,irn1,irn2,irn3}. We here propose a strategy to deal with missing values while using the \emph{fluctuation of similarity method}. To generate a test time series, we consider the same R\"{o}ssler model as introduced in the previous section and randomly remove some of the values in the time series. The amount of time points removed from the time series  are given in terms of percentage of missing values. A straightforward application of the fluctuation of similarity  method will not work in such a case, due to the incompatibility of embedding a time series in delayed coordinates with missing values. The first step of our strategy for dealing with treatment of missing values involves replacing the missing values with a flag (e.g. a NaN character). Then we continue to embed the time series in time delayed coordinates, with some of the coordinates just being the flags. But this would make the numerical calculation of a distance metric impossible, to get over this issue we recommend to use the Chebyshev distance, rather then euclidean distance as done all throughout this work. Chebyshev distance between two vectors ${\bf{x}}_{j}$  and ${\bf{x}}_{l}$ is given by  $\| {\bf{x}}_{j} - {\bf{x}}_{l} \|=\max\limits_i\{|x_{j}^i-x_{l}^i|\}$, where 
$x_{j}^i$ is the $i$th component of the vector ${\bf{x}}_{j}$. It ignores the non-numerical flags and maximum is only calculated over the numerical values. Thus, it returns non-numerical values only in the rare case when all the components of both the vectors are non-numerical flags.   Whereas  euclidean distance cannot be calculated even if there is a single non-numerical flag present, which is the most common occurrence when we have missing values. Therefore, we prefer using Chebyshev distance over euclidean distance. Using the same delay and embedding dimensions as in the previous section, we present the result at different amounts of missing values in Fig.~\ref{fig:miss} (c-e). The error bar on the values of $\sigma_S$ were obtained from $1000$ different realizations of missing values. The solid horizontal red lines are again the mean significance levels for different realizations of missing values.  In  Fig.~\ref{fig:miss} we observe that the above strategy seems to work to certain amount of missing values in the data. The embedding parameter used in this case were exactly the same as used in previous example.        
 
\subsection{A note about embedding parameters and window size} \label{sec:stat-empar}

In the examples above, we have used a rather high embedding dimension, which is due to the
fact that the systems we are considering have one of its parameters
varying with time (such as Bakers' map and Lorenz system with a drift and R\"ossler system with  nonlinear transitions). This converts the systems
into non-autonomous systems. Taken's theorem is not valid for such a
system. Hence, we cannot take the embedding dimension $2m+1$ as prescribed
by the Taken's theorem ($m$ is the known dimension of the system) \cite{phar_3,phar_2}. Though, there is no specific embedding theorem
for such systems but heuristic arguments in~\cite{hegger} state that a proper
choice for the embedding dimension should be larger than $2(m+P)$ where $P$ is
the number of time varying parameters of the system. It has been suggested that
this technique of \emph{``overembedding''} a time series helps in overcoming both
nonstationarity and noise effects \cite{hegger,verdes}. We will continue using high embedding dimension in the next section, where we would be applying our method in the analysis of crime record's time series, as these time series have originated from a system (society) which is not only high dimensional but also a large parameter space. So, $2(m+P)$ must be a
large number. In the crime record's time series used below,  apart from visible non-stationarity
the time series also has a high amount of noise which is also visible by eye and via its power
spectrum. Hence, a high embedding dimension is an appropriate choice.   

In Fig.~\ref{fig:conse} we have shown a quick convergence of
$\sigma_{S}$ on taking large enough window sizes, which in
turn gives the measure dependence on the structure of the attractor
through the effective dimension $D_F$. By taking overlapping windows,
we avoid reducing the amount of data appreciably. The presented
method differs in one very basic aspect from other methods, in particular those based on
recurrence properties. In many of them one first takes a window over
the data (or embedded vectors) and then calculate some measure based
on the recurrence property \cite{casdagli,norbert_phyrep,trans1,trans2,trans3}. This brings the relationship between
windowing, dimension and delay. In our method we follow a different
approach, first of all the recurrence distances of a point over the
whole time series are calculated and, then, by comparing each
consecutive time point, we calculate the measure $S_{j | j+1}$ for each
point. Till this step we have no windowing. In the next step we
calculate the fluctuations in this measure by taking windows. The way
we have defined the significance test, the window size now helps in
resolving time scales on which we wish to see the transitions. The
real task of windowing is to give control over resolving time scales
for transitions.

\section{Application to social dynamics}

Now we present an application of our method to an observed time series in social dynamics. Crime rates in society might be interpreted as following some nonlinear dynamics and affected by political, economic, and social situations \cite{nld-crm}. Analytic methods of time series analysis and agent based modeling have been used to predict and quantify the evolution of crime rates in different settings and societies \cite{ts1-crm,ts2-crm,ts3-crm,ag1-crm,ag2-crm}. Various methods from the rich paradigm of nonlinear time series analysis do not appear to have been applied to
 available data sets of crime records. We here analyze time series of robberies and homicides in the United States from  1975 to 1993 with monthly resolution. With this analysis we attempt to understand the nature of relationship if any between unemployment and robberies, and unemployment and homicides over this period \cite{uar1-crm,uar3-crm,ts2-crm}.

\subsection{Data source}
The source of data studies here on monthly robberies and monthly homicides is ICPSR (Inter-university Consortium for
Political and Social Research) study 6792 (Uniform Crime Reports: Monthly Weapon-Specific Crime and Arrest Time Series, 1975-1993). The source of unemployment data is the US Bureau of Labour Statistics (http://www.bls.gov/data/), using the monthly levels of unemployment for the whole US for the period 1975-1993. In Fig.~\ref{fig:crim1} (a) and  Fig.~\ref{fig:crim2} (a), black lines correspond to monthly robberies and monthly homicides respectively and blue dotted lines represent the unemployment rate over the same period. We have removed the linear trend from monthly robberies and monthly homicides time series by subtracting a linear least squares fit to the data.    

\subsection{Results}

\begin{figure}
 \centering
  \includegraphics [width=\columnwidth]{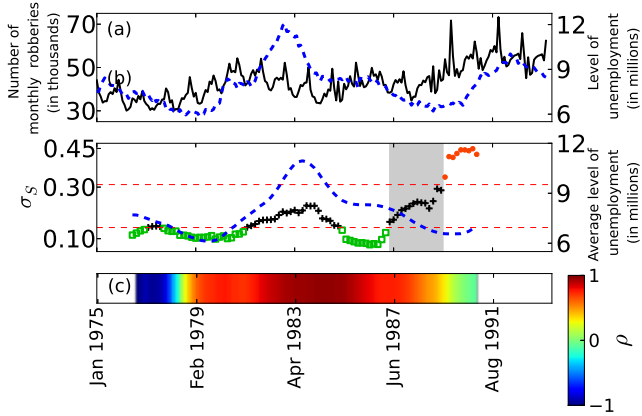}
  \caption{\small { (Color online) (a) Black line indicates the monthly robberies in the US between 1975 to 1993, whereas the blue dotted line is the monthly unemployment rate between the same period. (b) Values of $\sigma_S$ calculated for monthly robberies time series, represented by green open squares, black + signs and orange dots. The blue dotted line is the average values of monthly unemployment rate, calculated exactly with the same window sizes as used for $\sigma_S$.  Note the change in the values of $\sigma_S$ from green open squares to orange dots, representing a transition from one dynamical regime to other, as also highlighted with the grey Colored band. (c) Continuous color variation shows running windowed linear cross-correlation $\rho$ between  average values of monthly unemployment rate and values $\sigma_S$. Observe the high correlation between the two until 1987.}}
  \label{fig:crim1}
\end{figure}  \begin{figure}
 \centering
  \includegraphics [width=\columnwidth]{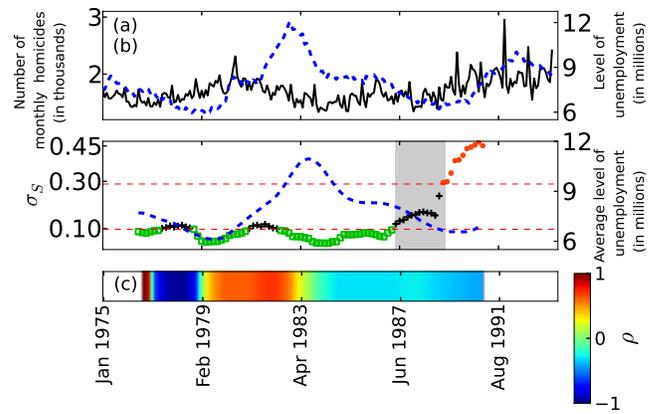}
  \caption{\small { (Color online) (a) Black line indicates the monthly homicides in the US between 1975 to 1993, whereas the blue dotted line is the monthly unemployment rate between the same period. (b) Values of $\sigma_S$ calculated for monthly homicides time series, represented by green open squares, black + signs and orange dots. The blue dotted line is the average values of monthly unemployment rate, calculated exactly with the same window sizes as used for $\sigma_S$.  Note the change in the values of $\sigma_S$ from green open squares to orange dots, representing a transition from one dynamical regime to other, as also highlighted with the grey Colored band. (c) Continuous color variation shows running windowed linear cross-correlation $\rho$ between  average values of monthly unemployment rate and values $\sigma_S$. Observe the low correlation between the two for almost over the whole of time period. }}
  \label{fig:crim2}
\end{figure}

The calculation of $\sigma_S$ for monthly robberies and homicides time series was done  using a window size of $20$ months with $90\%$ overlap, embedding dimension $12$, and delay of $3$, plotted in Fig.~\ref{fig:crim1}(b) and  Fig.~\ref{fig:crim2}(b). As emphasised in the discussion above, higher values of $\sigma_S$ correspond greater variability or complexity in the dynamics; while low values correspond to low complexity in the dynamics. For monthly robberies time series  in Fig.~\ref{fig:crim1}(b) we observe low values of $\sigma_S$ until 1982 (green open squares). Between 1983-1985 we also observe lower values of $\sigma_S$ but in a statistically insignificant regime (black plus signs). Then close to 1987 there is a significant increase in the values of $\sigma_S$ (orange dots) and the values cross the significance band during a transition between the period 1987--1990 (highlighted by a grey band). We uncover a similar transition in Fig.~\ref{fig:crim2}(b) occurring close to 1987  (see the grey band in both figures covering the period between 1987--1990.)

If we closely observe the original time series of robberies and homicides, then it is visible even to the naked eye that there are higher variabilities and larger fluctuations after this period. A fact to be noted here is that crimes in the US across all the categories of crime started to drop in the 1990's and this drop has continued since  \cite{drp1-crm,drp2-crm,drp3-crm,drp4-crm}. Several reasons have been hypothesized for this decrease, including increased incarceration \cite{drp5-crm}, more police \cite{drp6-crm}, the decline of crack use \cite{drp7-crm}, legalized abortion \cite{drp1-crm}, improvement in the quantity and quality of security \cite{drp3-crm} and changing demographics \cite{drp4-crm}. Our time series analysis above only brings forward the point that some fundamental change in the dynamics of crime in the US occurred in the late 1980's and early 1990's, leading to continuous drop in the crime rate in the following decades.  

In Fig.~\ref{fig:crim1} (c) and  Fig.~\ref{fig:crim2} (c), the continuous color variation gives the cross correlation $\rho$ between $\sigma_S$  and unemployment rate averaged over exactly the same time windows as $\sigma_S$. The blue curves in the middle panels of Fig.~\ref{fig:crim1} and  Fig.~\ref{fig:crim2} correspond to this averaged unemployment rate. In the case  of unemployment and robberies, we observe high positive values of cross correlation ($\rho \sim 1.0$) between the two curves from 1979 to 1989 and then an abrupt breakdown of this correlation, indicating  some fundamental shift in the crimes related to robberies around this time. In our second case of homicides, however we do not  observe any such relation between unemployment and homicides: the values of cross correlation between $\sigma_S$ and  average unemployment are rather low and fluctuating between negative and positive values. That is, the signals of unemployment rate driving variability and complexity in dynamics of robberies before the 1990's are quite apparent but they do not seem to play any significant role in homicides.     

\begin{figure}
 \centering
  \includegraphics [width=\columnwidth]{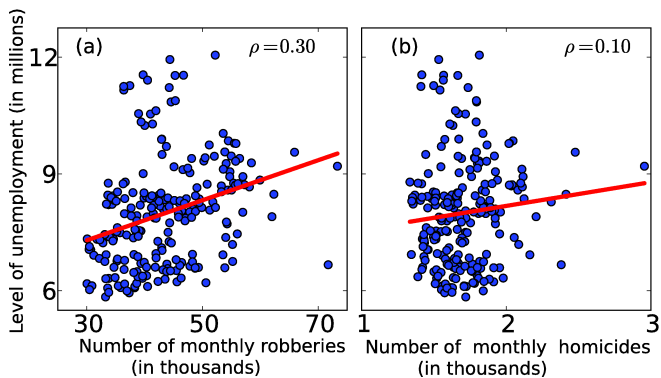}
  \caption{\small { (Color online) (a) Scatter plot between the level of unemployment and the monthly robberies. (b) Scatter plot between the level of unemployment and the monthly homicides. $\rho$ is the cross-correlation between the plotted variables and the red color line is the linear fit. Note the low correlations between the plotted variables in both (a) and (b).}}
  \label{fig:crim3}
\end{figure}

Sociologists have pointed out that the relationship between unemployment and robberies is a rather complex one: increasing unemployment increases the criminal motivation (unemployed individuals are more motivated to indulge in robbery for their financial needs and survival)  but it also decreases the criminal opportunity (more men start to stay at home, so less opportunity for criminals to break into homes), creating a counter balancing effect \cite{uar2-crm}. Hence, we cannot expect a linear relationship between both. We have also not observed a strong linear correlation (see Fig.~\ref{fig:crim3}) or a Granger causal relationship between these two variables. What our above analysis shows is that unemployment may have been driving the \emph{complexity or variability} in the  dynamics of robberies prior to the late 1980's and early 1990's. The breakdown in this relationship corresponds to a time period when the crime rate in the US started to steadily drop, due to several reasons  discussed in detail in the references \cite{drp1-crm,drp2-crm,drp3-crm,drp4-crm,drp5-crm,drp6-crm,drp7-crm}.     

An important perspective in criminology has been conflict theory, where it is considered that economic deprivations  influence crime rates \cite{ts3-crm}, but there does not exist a conclusive empirical support for this relationship \cite{uar1-crm, uar3-crm, uar4-crm}. In our analysis, if we treat the unemployment rate as being one of the economic indicators then we observe an episodic relationship between robberies and unemployment but the same cannot be said for homicides. Undoubtedly multiple interconnected factors including economic indicators drive crime rates. To accept or reject the economic deprivations perspective of crime, one would need to do extensive analysis of different social and economic indicators. As demonstrated above, our method could be useful in such analyzes and in other endeavours where similar questions could arise.    

\section{Conclusion} 
 
 Developing a set of methods that can be used to distinguish distinct dynamical regimes and transitions between them in a given time series has been a challenge in nonlinear time series analysis with wide applicability in a variety of fields. We have recently proposed a new method, based on computation of nonlinear similarities between time points of a univariate time series \cite{nmalik_epl1}.  The method is  robust, automatized, and computationally simple  and can be used even in cases with shorter time series, or missing values, or observational noise.  Here we have presented some new analytical findings, where we have related this measure to some classical concepts in nonlinear dynamics such as attractor dimensions and Lyapunov exponents. We have shown that the new measure has linear dependence on the variation of change in dimensionality or complexity of the attractor. Also, it measures the variance of the sum of the Lyapunov spectrum. One of the problems we have studied in detail with this method is identification of transitions in dynamics when the parameters of the system are also evolving with dynamics.  The proposed method is able to identify these most subtle of transitions, even including those  where such evolution of parameter induces only a drift or nonstationarity in the dynamics. Also,  employing a wide variety of prototypical model systems we have demonstrated the practical usefulness of this method. 
 
Furthermore, we have used this method to analyze a time series from  social dynamics, studying time series of US crime from 1975 to 1993. In doing so we have attempted to understand the nature of the relationship between crime rates (robbery and homicides) and unemployment levels during this period. We have found a dynamical transition in the late 1980's  in both homicide and robbery  rates and also found the dynamical complexity in robbery rates was driven by unemployment before this transition in 1990's.

\acknowledgments

N Malik  and P J Mucha acknowledge support from Award Number R21GM099493 from the National Institute of General Medical Sciences. The content is solely the responsibility of the authors and does not necessarily represent the official views of the National Institute of General Medical Sciences or the National Institutes of Health. Y Zou is supported by the National Natural Science Foundation of China (Grant Nos. 11305062, 11135001).  N Marwan and J Kurths are supported by the Potsdam Research Cluster for Georisk Analysis, Environmental Change and Sustainability (PROGRESS, support code 03IS2191B).


\end{document}